\documentclass[showkeys,prb,reprint]{revtex4-1}
\usepackage{hyperref}
\usepackage{graphicx}

\begin{document}


\title{Synchrotron radiation based photoemission studies of indium oxide passivation of ZnO(0001) surface at high temperature}

\author{Kumarappan Kumar}
\email[]{kumar.kumarappan2@mail.dcu.ie}
\author {Greg Hughes}
\affiliation{School of Physical Sciences and National Centre for Sensor Research, Dublin City University, Dublin-9, Ireland.}


\begin{abstract}
Passivation of surface reactive ZnO(0001)-Zn terminated surface is carried out by in-situ deposition of indium metal and post deposition annealing at high temperature (600~$^{\circ}$C - 1000~$^{\circ}$C). After the every cycle of indium deposition and annealing the ZnO(0001) surface is characterised by synchrotron radiation based photoemission. The photoemission studies confirmed the formation of deposited indium turned into indium oxide (In$_{2}$O$_{3}$ and In$_{2}$O$_{X}$) as surface passivation layer. The core level spectra for Zn, C, O, were acquired at different photon energy for high surface sensitivity. Overall core level peaks (Zn 3p, In 4d, O 1s) having average binding shift ranging from 0.5 eV to 0.8 eV. The valance band spectra shows fermi level shift of 0.9 eV after passivation at 1000~$^{\circ}$C and work function raises form 4.06 eV to 4.77 eV. Downward band bending of ZnO(0001) surface is turned into upward bending of 0.98 eV by indium oxide passivation at high temperature, which is suitable for making ZnO schottky contact and devices.   

\end{abstract}


\maketitle

\section{Introduction}

It is well known the ZnO surface have plenty of surface contaminations such as hydrocarbon, hydroxide and water were reported elsewhere~\cite{Coppa2003}. To remove those contaminations from ZnO surface both the in-situ and ex-situ surface cleaning methods (1) wet chemical cleaning~\cite{Kumar2008} (2) atomic hydrogen~\cite{Kumar2015} $\&$ oxygen cleaning were employed and their results were presented in previous studies. By all the surface cleaning methods employed, it was found only the carbon contamination can be removed completely, whereas the hydroxide contamination only get reduced by small amount and stays on the surface without any change. Due to the residual hydroxide on the ZnO surface had hampered the attainment of good rectify behaviour or making reliable schotty contact~\cite{Brillson2011}. As an alternative approach to surface cleaning in this work the plan is to employ the surface passivation methods to achieve a good rectify behaviour on ZnO surface.

To maintain the chemical, electronic, optical, thermal and ambient stability of the semiconductors surface the passivation were carried out on the variety of semiconductors. In detail, the semiconductor surface passivation was done for the following reasons (a) to saturate the dangling bonds (b) to stabilise the chemical bond under atmospheric pressure and operating temperatures (c) to prevent the diffusion both inward and outward of the surface (d) to control the excess of charge on surface and to produce the controllable surface band bending~\cite{Dannykim2001}. In the same time the surface passivation layers (a) should not damage the surface, (b) not to introduce further states into forbidden gap, (c) not to create the strain-induced defects on surface, etc~\cite{Dannykim2001}. 

Surface passivation of ZnO was carried out on bulk single crystals to nanostructures for different device applications such as making Schotty contacts~\cite{Skim2004}, enhance electrical conductivity~\cite{Cclin2005}, improve the luminescence~\cite{Tsekiguchi1997}, and enhance the gas sensing property~\cite{Gqi2014}, etc. Most of the ZnO surface passivation were done by chemical methods (just treating in solvents) than physical methods (surface processing in UHV environments). For making the reliable schottky contacts the oxygen passivation of ZnO surface was widely used to create the upward band bending of $\approx$0.5 eV. This oxygen passivation of ZnO was done by various ways, treating in hydrogen peroxide solutions~\cite{Rschifano2007}, pulsed laser excited oxygen~\cite{Msoh2007}, processing in remote oxygen~\cite{Bjcoppa2004} / ozone plasma~\cite{Kip2004} etc. Sulphur passivation were also employed to making rectifying contacts by treating with (NH$_{4}$)$_{2}$S$_{x}$ and got upward band bending of 0.71 eV~\cite{Yjlina2006, Yjlin2009}. Further different types of dielectric passivation were also carried out for developing reliable of schotty contacts such as HfO$_{2}$~\cite{Takrajewski2011}, CaHfO$_{2}$~\cite{Hvon2010}, etc. As generally known, the metal deposited on ZnO mostly turned into as Ohmic contact. An alternative approach of deposition metal oxide on ZnO and fabricating reproducible Schottky contacts were recently reported. The silver, copper, iridium and platinum oxide based ZnO schotty contacts showed high rectifying performance with effective barrier height above 1.0 eV~\cite{Mwallen2007, Martin2008} which is better than metal schottky contacts. Hence metal oxide passivation as well as metal oxide contacts on ZnO had made a significant improvement in fabrication of good performing schotty contacts.

In this study, ZnO(0001)-Zn terminated surface was chosen based on the results from previous studies~\cite{Kumar2015} , since it is the toughest surface to clean the contaminations and to control the surface band bending. Instead of depositing indium oxide directly on ZnO surface along with contamination surface. A two-step process was employed (1) deposition of pure indium metal on ZnO surface (2) post deposition annealing at high temperatures (\textgreater500~$^{\circ}$C). By this method, the deposited indium metal will turn into indium oxide by two possible ways either by natural oxidation in the UHV atmosphere or react with oxygen atoms from hydroxide / zinc oxide. The aim of this study is to explore this formation of the indium oxide at high temperature and its interaction with zinc oxide. As the result of it how the Fermi level position or surface band bending, work function of ZnO(0001) surface is changes and whether it is possible to create upward band bending by this process.


\section{Experimental}

ZnO single crystal of size (5 mm x 5 mm x 0.5 mm) which is polished on ZnO(0001)-Zn terminated surface was purchased from Crystal GmbH, Germany. The crystal was freshly opened from pack and loaded into the vacuum chamber without any pre-surface cleaning in solvents. Hence the sample surface must have the adsorbed adventious carbon and moisture/ water etc. So the sample was left in the load lock of pressure 2 x 10$^{-5}$ mbar about 10 hours to get remove the volatile contaminants. Then the sample was transferred to analysing chamber of pressure 4.4 x 10$^{-10}$ mbar. After recording the photoemission spectra for as received condition and annealing at 400~$^{\circ}$C was carried out to remove water, hydrocarbons contaminants, again photoemission spectra were acquired. The deposition of indium was done by resistive heating of tantalum pouch which is filled with tiny indium metal pieces and the evaporation temperature is between 350~$^{\circ}$C - 400~$^{\circ}$C under the pressure of \textless  1 x 10$^{-7}$ mbar. The duration of deposition is about 30 minutes and post-deposition annealing was performed after every deposition cycle by successive increase in temperatures. The post deposition-annealing temperature started from 600~$^{\circ}$C to 1000~$^{\circ}$C about 30 minutes in steps of 100~$^{\circ}$C. After every annealing the sample was cool down to the room temperature (30~$^{\circ}$C) and photoemission spectra were recorded. These photoemission experiments were carried out on the SX700 beamline at Institute for Storage Ring (ISA) in the University of Aarhus, Denmark in an UHV system (2 x 10$^{-10}$ mbar) with a typical photon flux of 1x10$^{10}$ photon/sec. The system was equipped with hemispherical (VG-CLAM2) electron energy analyser which was operated at pass energy of 30 eV. 

The photoemission energies were carefully chosen to acquire information samples top surface. The survey was acquired at 600 eV, then for core levels Zn 3d $\&$ In 4d (60 eV), Zn 3p (300 eV), C1s (345 eV), O1s (600 eV). The valence band spectra obtained at 60 eV and work function measurement was recorded at 40 eV with 20 eV pass energy. Peak fitting was done by both combination gauss $\&$ Lorentzian by using AA analyser~\cite{Aherrera2002}. 


\section{Results $\&$ Discussion}

\subsection{Core Level spectra}

\begin{figure*}[t!]
\includegraphics[width=1.65\columnwidth]{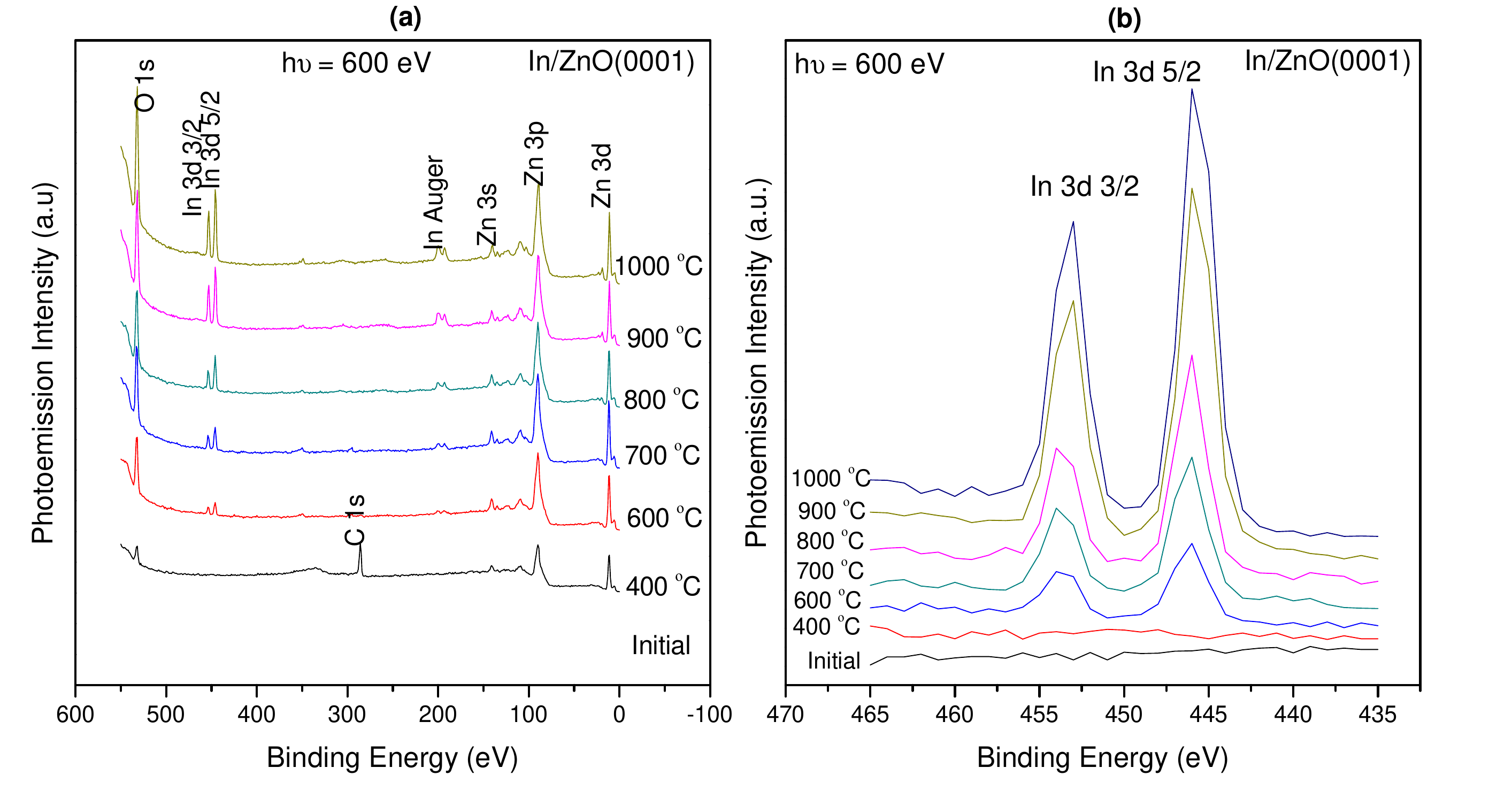}
\caption{(a) Survey scans for indium deposition on ZnO(0001) surface at different post annealing temperatures and (b) broad view of In 3d peak from survey scans\label{Fig1}}
\end{figure*}

The range of survey spectra for ZnO(0001) surface before and after indium deposition with respect to post deposition annealing temperature were shown Fig.\ref{Fig1}(a). In initial, (as received condition) only high intense carbon peak was seen and it implies the large amount carbon contamination is on the surface. After UHV annealing at 400~$^{\circ}$C, the carbon got reduced with emergence of zinc (Zn 3d, Zn 3p) and oxygen (O 1s) peaks. For the first cycle of indium deposition and post deposition annealing at 600~$^{\circ}$C suppressed the carbon peak with indium coverage, along with prominent peak shape of zinc and oxygen peak for ZnO were seen. Further increase in indium deposition cycle along with systematic increase in post deposition annealing temperature (700~$^{\circ}$C - 1000~$^{\circ}$C) had increased the intensity of all zinc, oxygen and indium peaks. In addition to core level peaks, indium auger peaks were also found. Since the kinetic energy of those peaks are at 402.6 eV and 410.6 eV.

The binding energy of the peaks in survey spectra as follows: O 2p - 6 $\pm$ 0.3 eV, Zn 3d - 10.75 $\pm$ 0.25 eV, Zn 3p - 89.75 $\pm$ 0.25 eV, Zn 3s - 141 $\pm$ 0.05 eV, In Auger - 193.25 $\pm$ 0.25 eV, 199.75 $\pm$ 0.25 eV, C 1s - 286.5 $\pm$ 0.5 eV, In 3d 5/2 - 446 $\pm$ 0.05 eV, In 3d 3/2 - 453.5 $\pm$ 0.5 eV, O 1s - 531.25 $\pm$ 0.25 eV. The In 3d peak taken from survey spectra is separately shown in Fig.\ref{Fig1}(b) which have the binding energy of the In 3d 3/2 peak stably at 446 eV. This indicates that the deposited indium metal on ZnO surface is transformed into indium oxide. Since binding energy of In 3d 3/2 - 446 eV (above 445.2 eV) is meant for complete oxidation of In into In$_{2}$O$_{3}$\cite{Gemcguire1973,Tmori2013}.

\begin{figure}[t!]
\includegraphics[width=1\columnwidth]{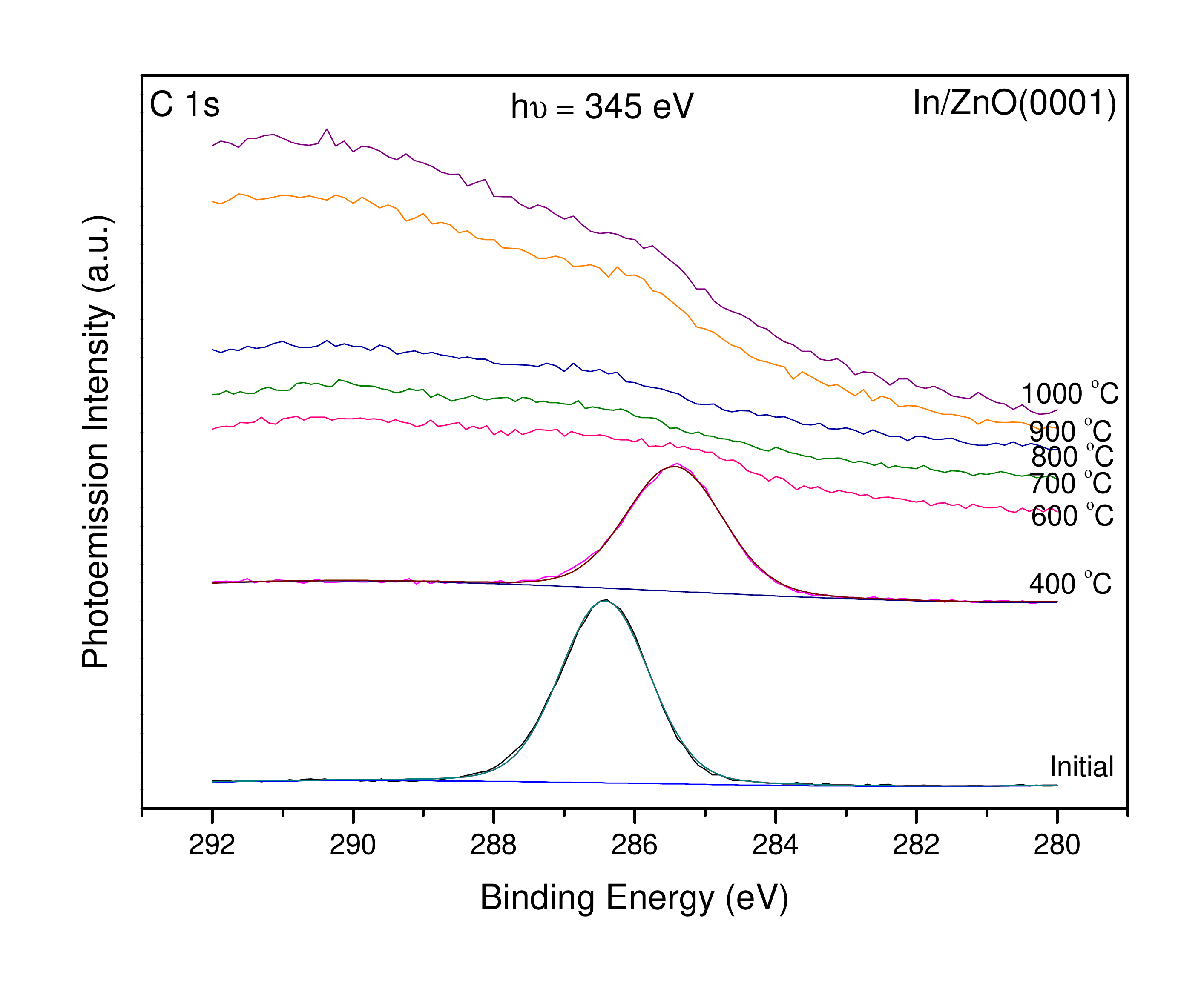} 
\caption{C 1s - carbon core level spectra of ZnO(0001) surface for different  cycles of indium deposition and post deposition annealing temperatures.\label{Fig2}}
\end{figure}

The narrow scans of carbon C 1s core level spectra obtained for photoemission energy 345 eV is illustrated in the Fig.\ref{Fig2}. Only the carbon peak was seen in as received condition and 400~$^{\circ}$C UHV annealing conditions. For further indium deposition and post annealing condition no more significant carbon peak is recorded. At 400~$^{\circ}$C, UHV annealing induced chemical shift of 1 eV (286.42 eV - 285.42 eV) is seen. The FWHM of the peak fit with gauss - 1.4 eV and lorentzian - 0.2 eV.

\begin{figure*}[t!]
\includegraphics[width=1.65\columnwidth]{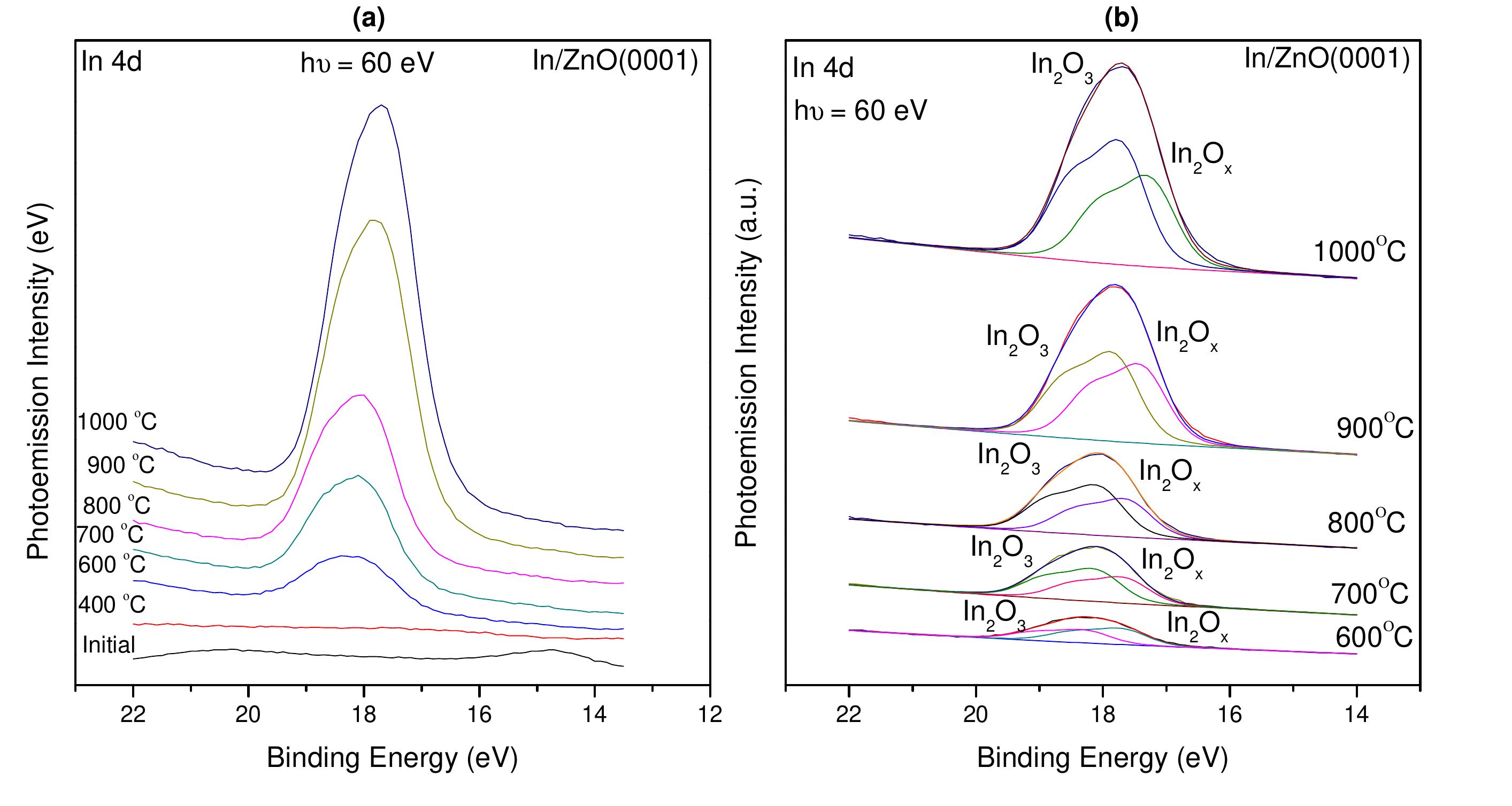}
\caption{In 4d - Indium core level spectra of ZnO(0001) surface for different indium deposition and post deposition annealing temperatures (a) raw spectra (b) peak fitted.\label{Fig3}}
\end{figure*}

The indium (In 4d) spectra acquired at the photoemission energy of 60 eV were illustrated in Fig.\ref{Fig3}. The comparison of raw indium spectra (Fig.\ref{Fig3}(a)) for different post deposition annealing temperatures shows the successive rise in intensity of the peak. This confirms the systematic growth of indium atomic layer on ZnO surface. The binding energy of the peak also moves from high (18.21 eV) to low (17.7 eV) with respect to annealing temperature is due to Fermi level movement (Fig.\ref{Fig3}(b)). These binding energy values were corresponds to the indium oxide. It is well known most of the metals (Al, Mg, Ti, V, Cu etc) including indium were readily get oxidised in ultra-high vacuum (UHV) atmosphere~\cite{Vehenrich1996}. Further by peak fitting (Fig.\ref{Fig3}(b)), two peaks were fitted for In 4d one peak for In$_{2}$O$_{x}$ - 17.46 $\pm$ 0.22 eV and other peak for In$_{2}$O$_{3}$ - 17.98 $\pm$ 0.29 eV. These binding energy values were consistent with previous studies reported\cite{Bbrennan2010}. The peak fit parameters of In 4d doublet peak are FWHM - 0.8 eV (gauss), 0.2 eV (lorentzian), splitting - 0.8 eV, and ratio - 0.66 eV.

\begin{figure}[t!]
\includegraphics[width=1\columnwidth]{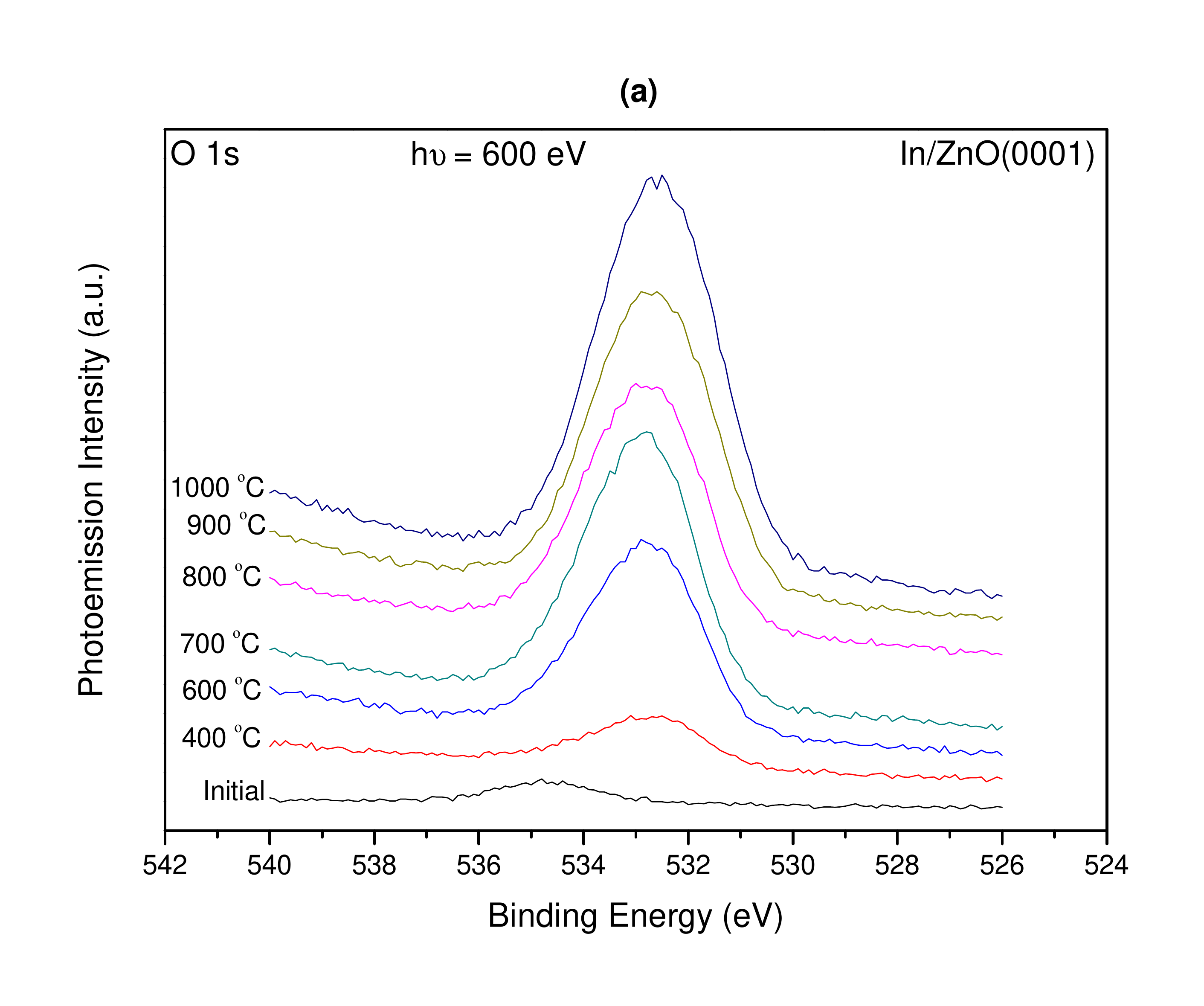}
\caption{Comparison of oxygen (O 1s) core level spectra of ZnO(0001) surface for different indium deposition and annealing temperatures.\label{Fig4}}
\end{figure}

\begin{figure}[t!]
\includegraphics[width=0.80\columnwidth]{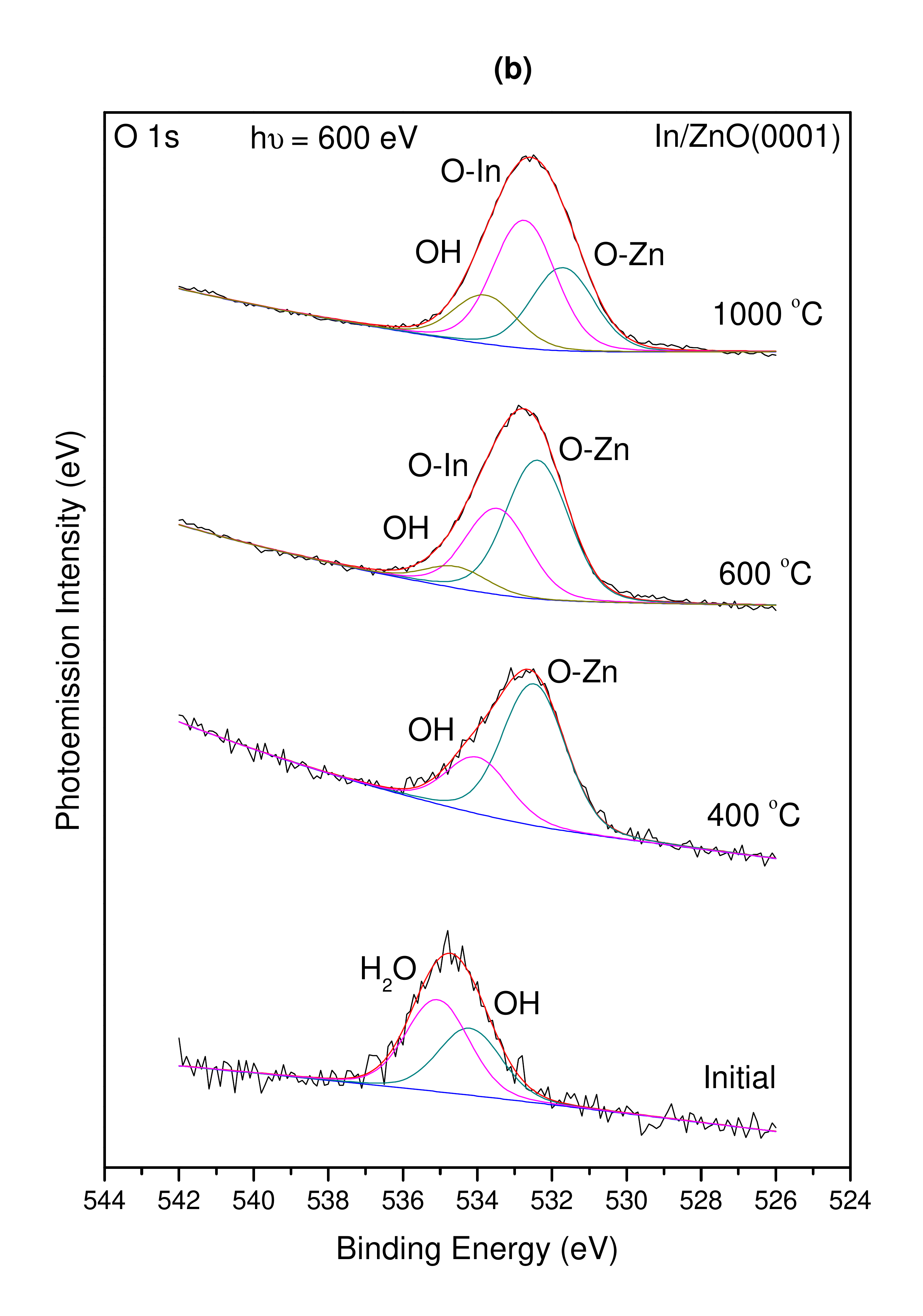}
\caption{Peak fitted Oxygen (O 1s) core level spectra of ZnO(0001) surface for different indium deposition and annealing temperatures.\label{Fig5}}
\end{figure}

The comparison of oxygen O 1s core level spectra for indium deposition cum different annealing temperature was illustrated in Fig.\ref{Fig4}. For initial condition, very low intensity peak is obtained at higher binding energy position (534.69 eV) is due to the large amount of carbon and water contamination on surface. This binding energy position 534.69 eV is meant for water. After removing the water on surface by annealing at 400~$^{\circ}$C the peak moves to lower binding energy 532.58 eV which meant for ZnO. Later by depositing indium and annealing at higher temperatures the oxygen peak stabilised at 532.75 eV. The oxygen O 1s peak were peak fitted (Fig.\ref{Fig5}) with 2 peaks for initial, 400~$^{\circ}$C annealing condition and 3 peaks for all indium deposited annealing (600~$^{\circ}$C - 1000~$^{\circ}$C) conditions. In initial condition the fitted peaks were for OH - 534.19 eV and H$_{2}$O - 535.06 eV and at 400~$^{\circ}$C UHV annealing the peaks were for O-Zn - 532.47 and OH - 533.99 eV. It is clearly seen the water was dominant in initial condition and at 400~$^{\circ}$C UHV annealing zinc oxide was dominant. For all indium deposited and annealing conditions(600~$^{\circ}$C, 700~$^{\circ}$C, 800~$^{\circ}$C) the peaks were at O-Zn - 532.4 eV, O-In - 533.4 eV and H$_{2}$O - 534.5 eV. Exactly a 0.4 eV high binding energy shift (O-Zn - 532.7 eV, O-In - 533.7 eV and H$_{2}$O - 534.8 eV) for 900~$^{\circ}$C and 1000~$^{\circ}$C is found because of Fermi level movement. After every indium deposition and increase in annealing temperature the indium oxide peak got increased without much change in zinc oxide and hydroxide peak was found (by comparing 600~$^{\circ}$C  to 1000~$^{\circ}$C). The FWHM for O 1s peak fit is 1.8 eV (gauss) and 0.3 eV (Lorentzian).

\begin{figure}[t!]
\includegraphics[width=1\columnwidth]{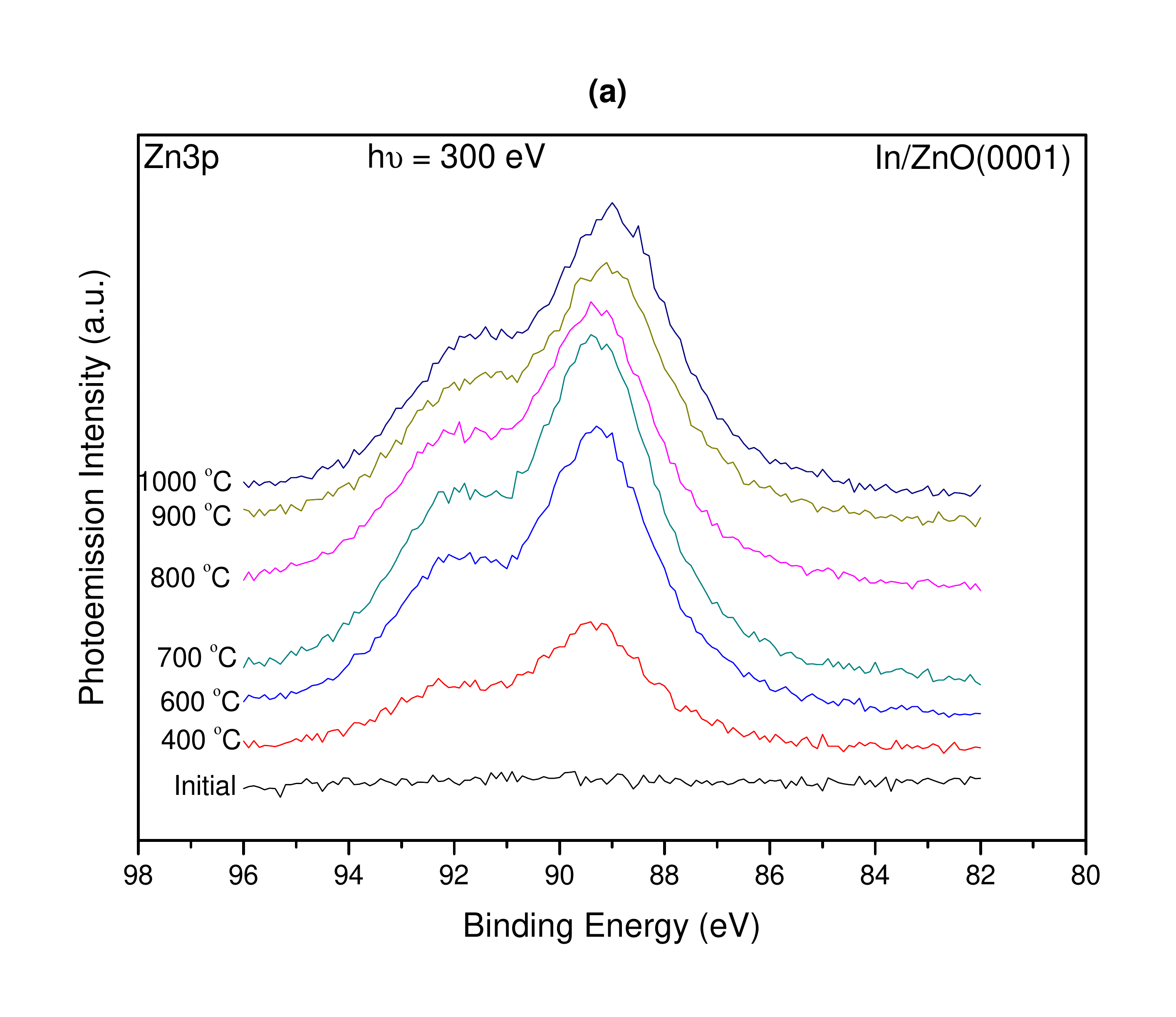}
\caption{Zn 3p - Zinc core level spectra of ZnO(0001) surface for different indium deposition and annealing temperatures (raw spectra).\label{Fig6}}
\end{figure}

\begin{figure}[t!]
\includegraphics[width=0.80\columnwidth]{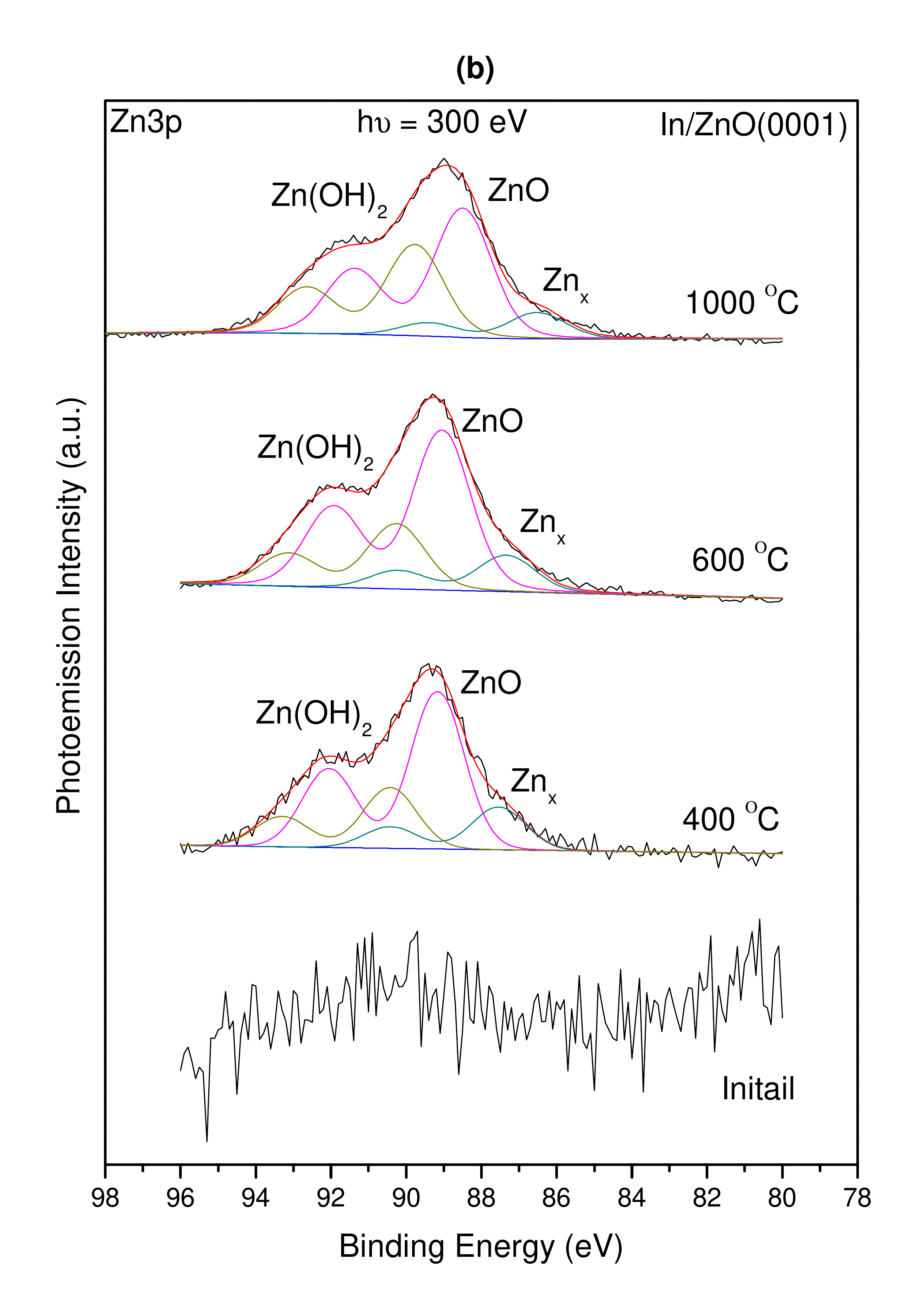}
\caption{Zn 3p - Zinc core level spectra of ZnO(0001) surface for different indium deposition and annealing temperatures (peak fitted).\label{Fig7}}
\end{figure}

At the photoemission energy of 300 eV the Zn 3p zinc spectra were obtained for indium deposition and different annealing temperatures is illustrated in Fig.\ref{Fig6} and \ref{Fig7}. The comparison of as recorded Zn 3p spectra was shown in Fig.\ref{Fig6}. For initial condition there was no more peak observed because complete coverage of contaminants on surface. For 400~$^{\circ}$C UHV annealing a signature Zn 3p spectrum for ZnO is seen. On further increasing on the indium depositions and annealing temperatures, for 600~$^{\circ}$C and 700~$^{\circ}$C there is increase in the intensity of Zn 3p peak, then afterwards (800~$^{\circ}$C, 900~$^{\circ}$C and 1000~$^{\circ}$C) peak intensity drops down due to the growth of indium oxide. The usual three peaks were fitted~\cite{Kumar2015} for Zn 3p spectra (Fig.\ref{Fig7}), surface state peak Zn$_{x}$ - 87.3 eV, oxide peak ZnO - 89.1 eV and hydroxide peak Zn(OH)$_{2}$ - 90.4 eV. For 900~$^{\circ}$C and 1000~$^{\circ}$C all three peaks shift toward lower binding energy about 0.5 - 0.8 eV. This binding energy shift is as result of Fermi level movement on ZnO surface. The FWHM of the peak fits are gauss - 1.6 eV and Lorentzian - 0.3 eV.

\begin{figure*}[t!]
\includegraphics[width=1.80\columnwidth]{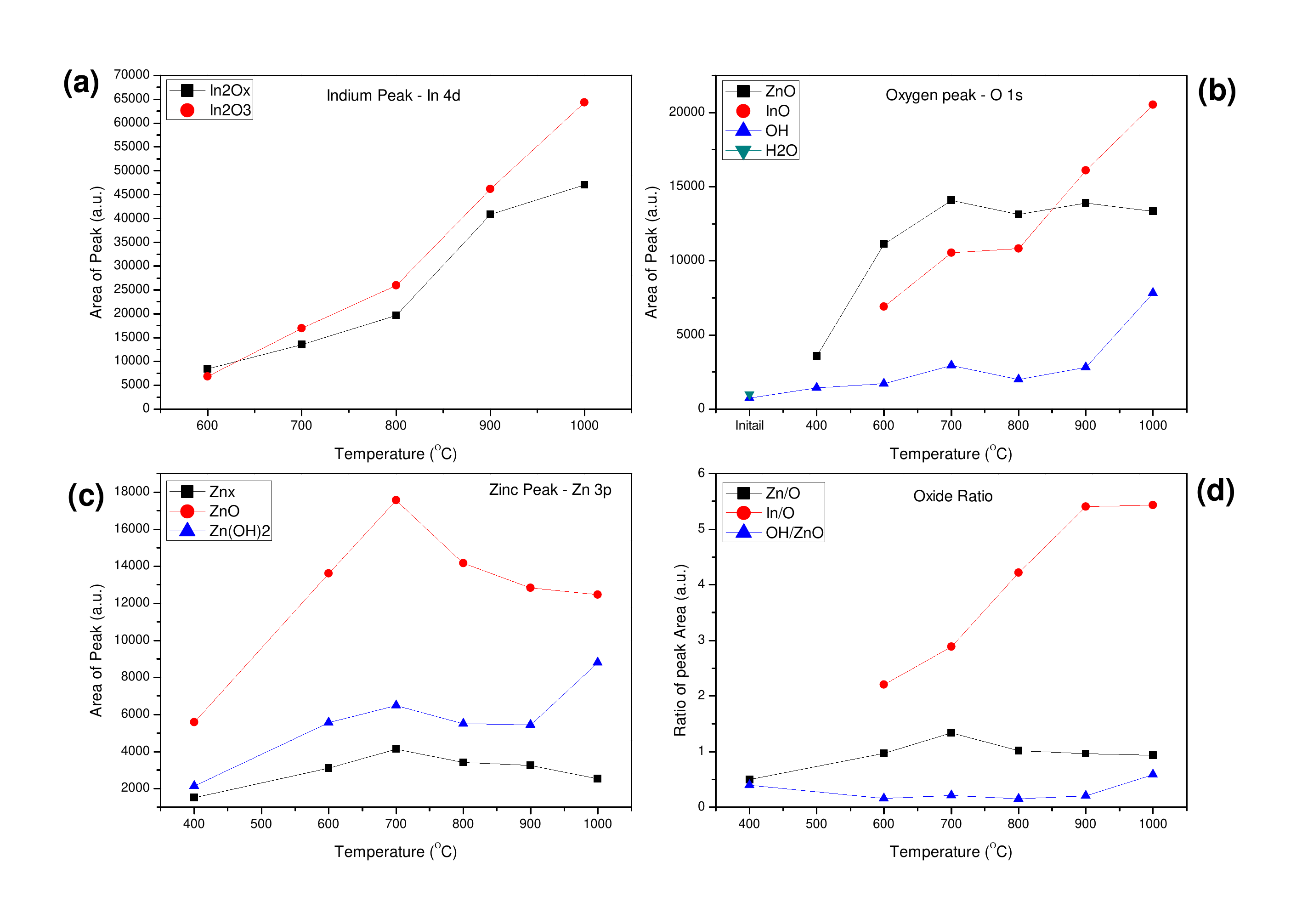}
\caption{Analysis of areas of peak fitted for (a) indium, (b) oxygen, (c) zinc, and (d) stoichiometry ratio of In/O, Zn/O, contamination ratio of OH/ZnO for different indium deposition with annealing temperatures.\label{Fig8}}
\end{figure*}

The peak area analysis was done to estimate the chemical composition variation on ZnO (0001) surface as effect of indium oxide passivation. Peak fitted area of core level peaks of In, Zn, O were separately plotted in the Fig.\ref{Fig8}. In the graph (Fig.\ref{Fig8}(a)) for indium peak (In 4d) both the completely oxidised indium (In$_{2}$O$_{3}$) and partially oxidised indium (In$_{2}$Ox) were in raising order. This shows there is systematic growth of indium oxide. In oxygen peak (O 1s) graph (Fig.\ref{Fig8}(b)) also similarly the oxygen bonded to indium (InO) is in rising order again confirms the uniform growth of indium oxide. Oxygen bonded to zinc (ZnO) in O 1s graph shows the rise in peak area up to 700~$^{\circ}$C and afterwards almost stay constant for rest of all annealing temperatures. This implies the removal of contaminations happened up to 700~$^{\circ}$C, so ZnO peak get increased and after no more impact on oxygen in ZnO due to indium oxide growth. 

The hydroxide (OH) in O 1s peak also shows similar trend as like in InO peak, but the peak area increase in small increments. In zinc Zn 3p graph (Fig.\ref{Fig8}(c)), the ZnO peak area increase up to 700~$^{\circ}$C (as like in ZnO in O 1s) and  for 800~$^{\circ}$C - 1000~$^{\circ}$C drops down systematically. So the increase in In 4d peak exactly affects the Zn 3p peak. Then the zinc hydroxide (Zn(OH)$_{2}$) peak also follows the same trend of hydroxide(OH) in O 1s peak. In the final graph (Fig.\ref{Fig8}(d)) for metal to oxygen ratio, In/O ratio is in rising trend, Zn/O ratio and OH/ZnO were in almost uniform. This oxide ratio graph indicates the no more change in ZnO and its hydroxide with growth of indium oxide.

\subsection{Valence band spectra}

\begin{figure*}[t!]
\includegraphics[width=1.65\columnwidth]{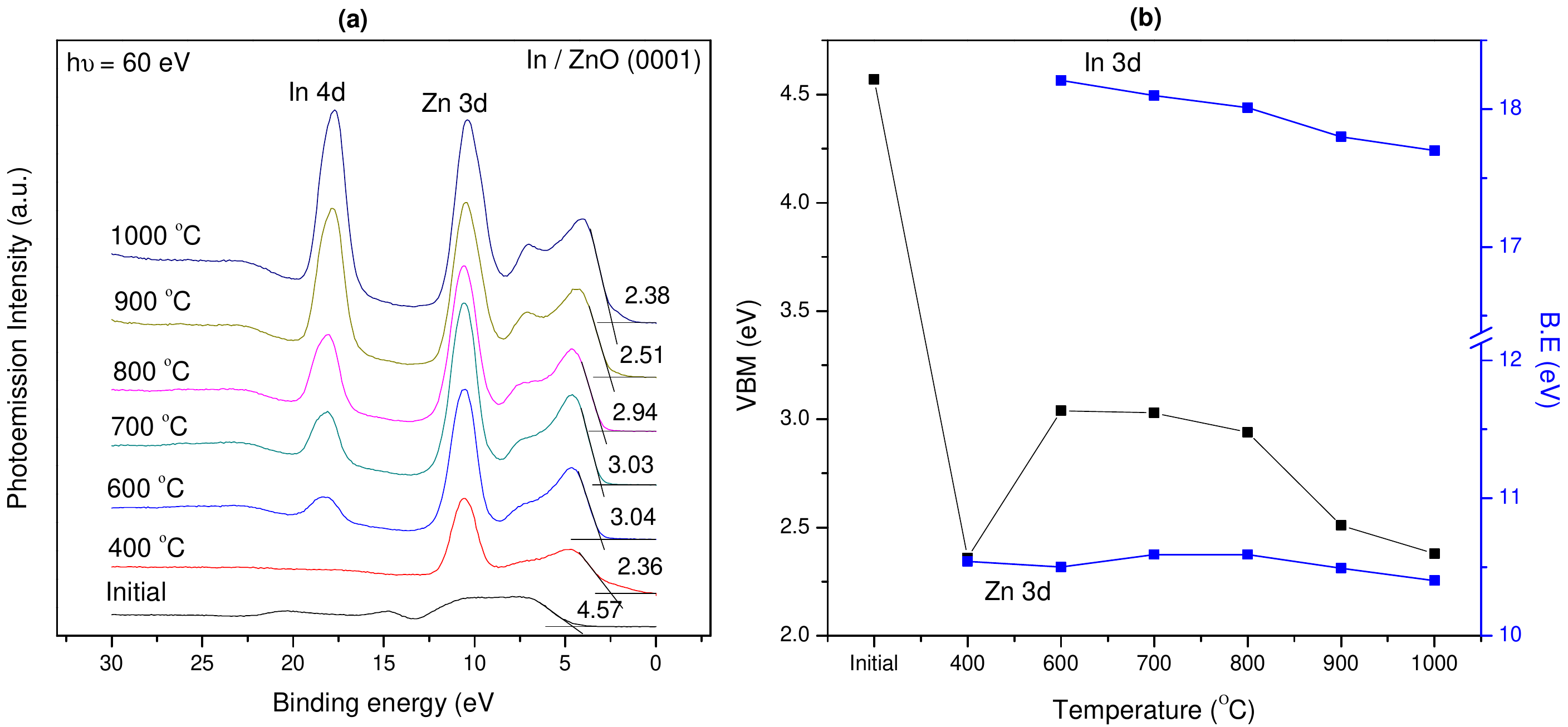}
\caption{(a) Valence band spectra for ZnO(0001) under different indium deposition cum annealing temperatures and (b) change in valence band maximum (VBM), Zn 3d Binding energy and In 4d Binding energy with respect to different annealing temperature. \label{Fig9}}
\end{figure*}

The Zn 3d and In 3d peaks containing valence band spectra for ZnO(0001) surface for different post deposition annealing temperatures were illustrated in Fig.\ref{Fig9}(a). No more peaks was observed in as received (initial) condition and Zn 3d peak get emerged after 400~$^{\circ}$C UHV annealing. After first cycle of indium deposition and annealing (600~$^{\circ}$C) a signature valence band spectra for ZnO(0001) surface is got with In 4d peak. On further increasing indium deposition and annealing temperatures had increased the intensity of In 4d peak along with Zn 3d peak too. The valence band maximum (VBM) the distance between Fermi level and valence band in semiconductor energy band was measured here from the valence band spectra. The VBM is measured by extrapolating the shoulder of valence band in lower binding energy region to background of the spectrum (bottom line). The variation of VBM value, Zn 3d and In 4d binding energy values were plotted in the graph (Fig.\ref{Fig9}(b)). The VBM value of 4.57 eV measured for contaminated as received ZnO(0001) surface is due to large amount of water and carbon stays on the surface. After removing water and reducing carbon contamination by annealing at 400~$^{\circ}$C dramatically drops the VBM to 2.36 eV which is not stable value still contaminants present on the top surface. On deposition of indium and annealing at 600~$^{\circ}$C, 700~$^{\circ}$C, 800~$^{\circ}$C the VBM value almost retains the same level of 3.04 eV - 2.94 eV which is due to the formation indium oxide on surface. On further depositing indium and annealing at higher temperatures 900~$^{\circ}$C, 1000~$^{\circ}$C really strong layer of indium oxide forms on the ZnO surface. Hence the VBM value moves to 2.38 eV which is stable because indium oxide act as passivation layer on ZnO(0001) surface and all contaminants were suppressed. On analysing the binding energy variation of Zn 3d and In 4d from the graph, Zn 3d binding energy vary about 0.2 eV between 10.59 eV to 10.4 eV. But the variation of In 4d binding energy is 0.5 eV between 18.21 eV to 17.7 eV. This variation of In 4d binding energy exactly correlate with variation of VBM values. So the Fermi level movement happened on ZnO surface is due to the passivation of indium oxide on ZnO(0001) surface.

\begin{figure}[t!]
\includegraphics[width=1\columnwidth]{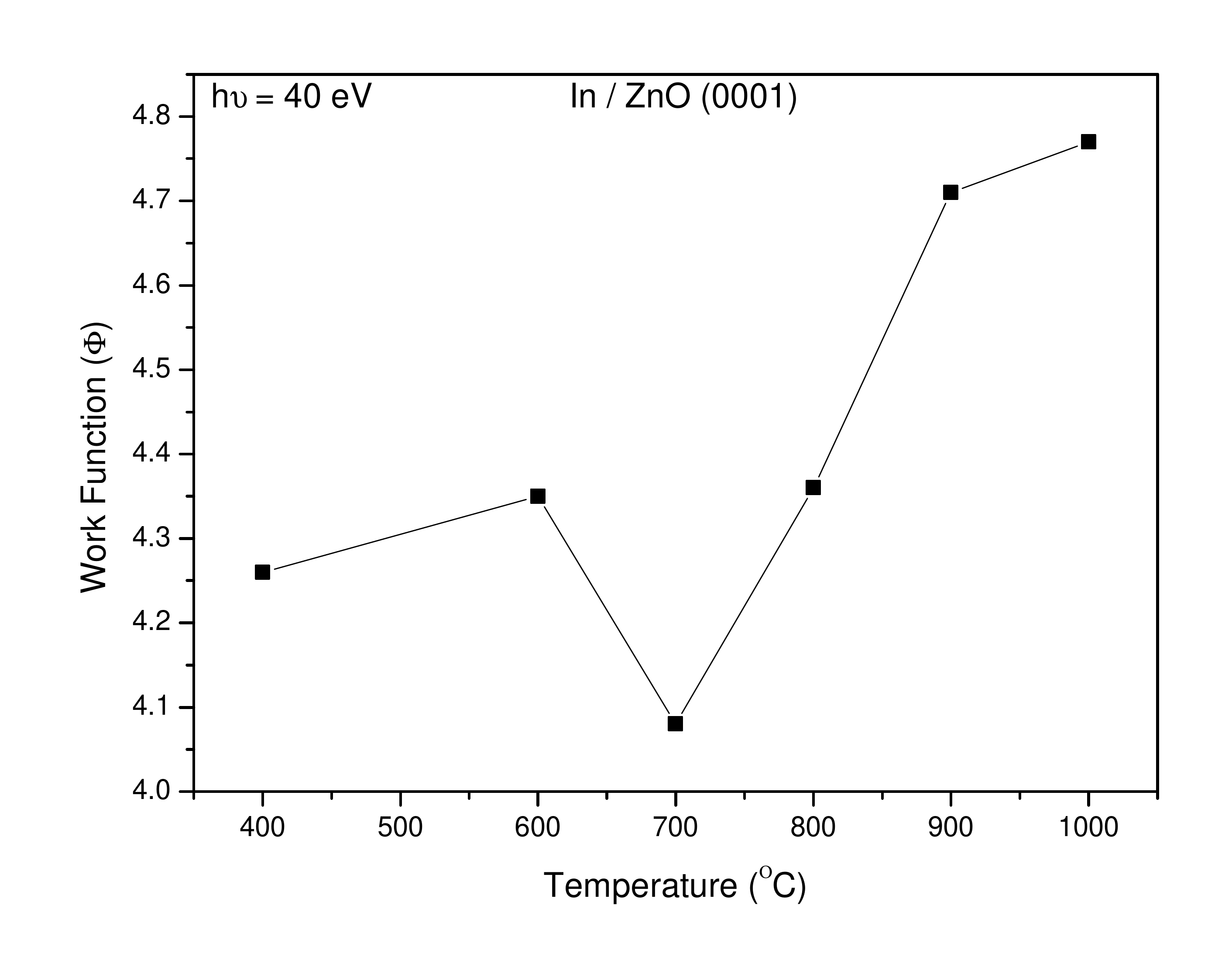}
\caption{Work function ($\Phi$) values obtained for indium deposition on ZnO(0001) surface for different annealing temperatures.\label{Fig10}}
\end{figure}

The work function values were separately measured by photoemission energy of 40 eV with 20 eV pass energy and plotted in Fig.\ref{Fig10}. For initial condition the work function value obtained is 4.06 eV which is generally meant for ZnO. On further annealing (400~$^{\circ}$C), indium deposition and annealing (600~$^{\circ}$C - 1000~$^{\circ}$C work function raises form 4.26 eV to 4.77 eV is due to the Fermi level movement exactly matches with variation of VBM values. But the drop in work function at 700~$^{\circ}$C is unclear. Over all, the analysis of work function values also confirms the Fermi level movement happened due to indium oxide passivation on ZnO (0001) surface.

\begin{figure}[hbt!]
\includegraphics[width=0.6\columnwidth]{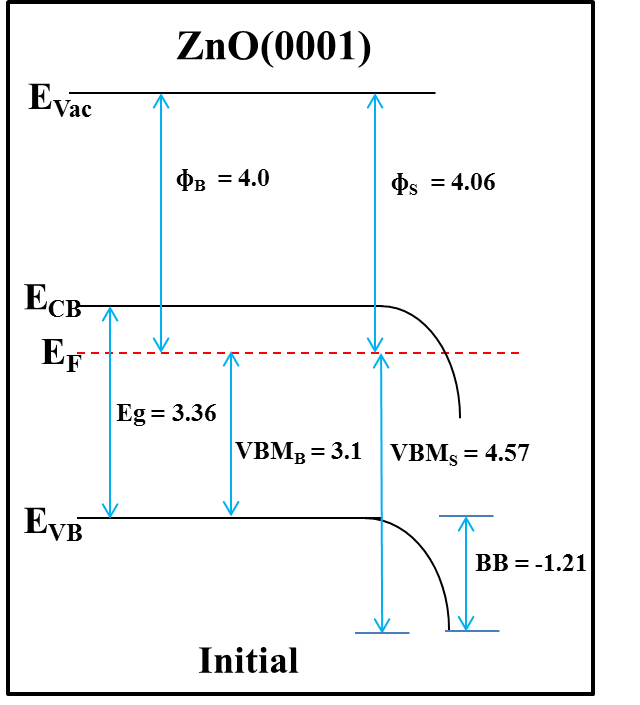} 
\includegraphics[width=0.6\columnwidth]{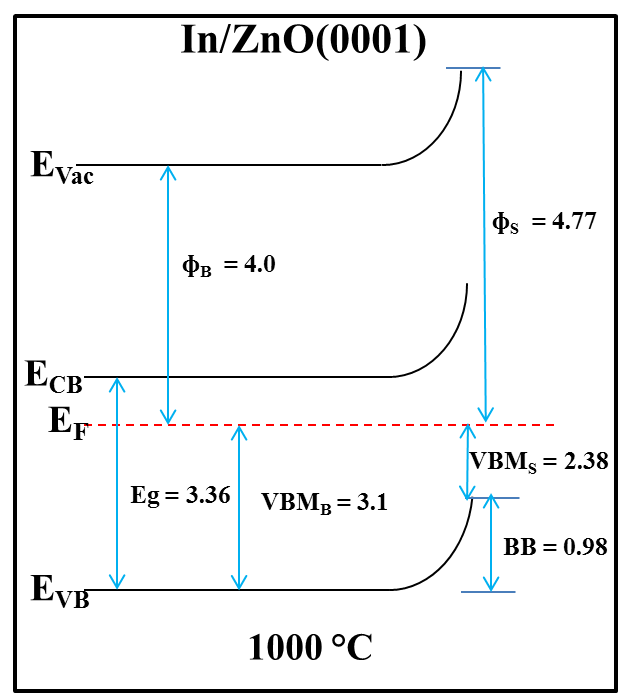}
\caption{Energy level band alignment of ZnO(0001) surface before and after indium deposition cum annealed at high temperatures.\label{Fig11}}
\end{figure}

The electronic structure of ZnO(0001) surface before and after indium deposition with selected annealing temperature was shown in Fig.\ref{Fig11}. The energy band gap of ZnO is E$_{g}$ = 3.36 eV, the Fermi level position (E$_{F}$), Work function ($\Phi_{B}$) and Valence band maximum (VBM$_{B}$) for Bulk ZnO is obtained from previous reports\cite{Kozawa2011,Rheinhold2013}. The values ($\Phi_{S}$, VBM$_{S}$) for ZnO surface were obtained from our experiment results here (Fig.\ref{Fig9} and \ref{Fig10}). The band bending values were calculated form difference between energy band gap and valence band maximum (BB = E$_{g}$ - VBM$_{S}$). In initial condition with high level of carbon and water contamination on ZnO(0001) surface leads to the downward band bending of -1.21 eV. This is because adsorbed water / OH on surface donate more free electrons to the surface and makes the surface n-type or ohmic\cite{Brillson2011}. After the water contaminations on surface were evaporated by UHV annealing at 400~$^{\circ}$C. The first cycle of indium deposition and annealing at 600~$^{\circ}$C form the first layer of indium oxide on surface.

This indium oxide layer neutralise available free electrons from OH and donate hole to the surface. As the result of it an upward band bending of 0.32 eV observed at 600~$^{\circ}$C. Further increasing the number of indium deposition cycles and annealing temperatures a good passivation layer of indium oxide was formed on ZnO surface at 1000~$^{\circ}$C. Hence ZnO(0001) exhibits the rectifying behaviour at 1000~$^{\circ}$C with upward band bending of 0.98 eV. These results shows that passivation of ZnO surface by metal oxide layer is better than employing the surface cleaning methods~\cite{Kumar2015} to create upward banding, which is suitable for making ZnO schotty contact and devices.   


\section{Conclusion}

As received ZnO(0001)-Zn terminated surface contains large amount carbon, water and hydroxide contaminates on surface were found as characterised by synchrotron radiation based photoemission spectroscopy. Annealing at 400~$^{\circ}$C removed all water on surface and ZnO photoemission peaks were detected. Along with carbon and hydroxide contaminations on ZnO surface a successful indium oxide layers formed by the process of depositing indium metal followed by annealing in UHV atmosphere at high temperatures (600~$^{\circ}$C - 1000~$^{\circ}$C). For every cycle of metal deposition and annealing, (indium oxide layer growth) a significant Fermi level movement away from conduction band was observed. In as received condition ZnO (0001) surface exhibit ohmic behaviour of downward band bending about -1.21 eV. After growth of indium oxide passivation layer on ZnO(0001) shows the rectifying behaviour with upward band bending of 0.98 eV. 

\begin{acknowledgements}

The authors wish to acknowledge financial support from Science Foundation Ireland under grant number (RFP Grant No. 06/RFP/PHY052). Access to the ASTRID synchrotron radiation source was funded under the EU Access to Research Infrastructure (ARI) programme and SFI Project (RFP 06/RFP/PHY035). We also acknowledge the assistance of ASTRID staff for support during the synchrotron radiation experiment. 

\end{acknowledgements}



\end{document}